\documentclass{article}

\usepackage{PRIMEarxiv}

\usepackage[utf8]{inputenc} % allow utf-8 input
\usepackage[T1]{fontenc}    % use 8-bit T1 fonts
\usepackage{hyperref}       % hyperlinks
\usepackage{url}            % simple URL typesetting
\usepackage{booktabs}       % professional-quality tables
\usepackage{amsfonts}       % blackboard math symbols
\usepackage{nicefrac}       % compact symbols for 1/2, etc.
\usepackage{microtype}      % microtypography
\usepackage{lipsum}
\usepackage{fancyhdr}       % header
\usepackage{graphicx}       % graphics
\graphicspath{{media/}}     % organize your images and other figures under media/ folder
\usepackage{pdflscape}

\title{A review of handcrafted and deep radiomics in neurological diseases: transitioning from oncology to clinical neuroimaging}

\author{ 
	\href{https://orcid.org/0000-0003-2751-790X}{\includegraphics[scale=0.06]{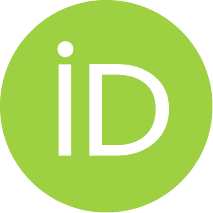}\hspace{1mm}Elizaveta Lavrova} \\
	Department of Precision Medicine\\
	GROW—School for Oncology\\
	Maastricht University, the Netherlands \\
	GIGA Cyclotrone Research Center\\
	University of Liege, Belgium\\
	\texttt{lavrovaliz@gmail.com} \\
	\And
	\href{https://orcid.org/0000-0001-7911-5123}{\includegraphics[scale=0.06]{orcid.pdf}\hspace{1mm}Henry Woodruff} \\
	Department of Precision Medicine\\
	GROW—School for Oncology\\
	Maastricht University, the Netherlands \\
	Department of Radiology and Nuclear Medicine \\
	Maastricht University Medical Centre, the Netherlands\\
	\texttt{h.woodruff@maastrichtuniversity.nl} \\
	\And
 \href{https://orcid.org/0000-0002-0398-4848}{\includegraphics[scale=0.06]{orcid.pdf}\hspace{1mm}Hamza Khan} \\
	Department of Precision Medicine\\
	GROW—School for Oncology\\
	Maastricht University, the Netherlands \\
	Biomedical Research Institute (BIOMED) \\
	University of Hasselt, Belgium \\
	\texttt{h.khan@maastrichtuniversity.nl} \\
	\And
 \href{https://orcid.org/0000-0003-2520-9241X}{\includegraphics[scale=0.06]{orcid.pdf}\hspace{1mm}Eric Salmon} \\
	GIGA Cyclotrone Research Center\\
	University of Liege, Belgium\\
        Neurology Department\\
        CHU Liège, Belgium \\
	\texttt{eric.salmon@uliege.be} \\
	\And
	\href{https://orcid.org/0000-0001-7961-0191}{\includegraphics[scale=0.06]{orcid.pdf}\hspace{1mm}Philippe Lambin} \\
	Department of Precision Medicine\\
	GROW—School for Oncology, \\
        Maastricht University, the Netherlands \\
	Department of Radiology and Nuclear Medicine \\
	Maastricht University Medical Centre, the Netherlands \\
	\texttt{philippe.lambin@maastrichtuniversity.nl} \\
 \And
 \href{https://orcid.org/0000-0002-4990-425X}{\includegraphics[scale=0.06]{orcid.pdf}\hspace{1mm}Christophe Phillips} \\
	GIGA Cyclotrone Research Center\\
        and GIGA In Silico Medicine \\
	University of Liege, Belgium\\
	\texttt{c.phillips@uliege.be} 
}

\begin{document}
\maketitle

\begin{abstract}
Medical imaging technologies have undergone extensive development, enabling non-invasive visualization of clinical information. The traditional review of medical images by clinicians remains subjective, time-consuming, and prone to human error. With the recent availability of medical imaging data, quantification have become important goals in the field. Radiomics, a methodology aimed at extracting quantitative information from imaging data, has emerged as a promising approach to uncover hidden biological information and support decision-making in clinical practice. This paper presents a review of the radiomic pipeline from the clinical neuroimaging perspective, providing a detailed overview of each step with practical advice. It discusses the application of handcrafted and deep radiomics in neuroimaging, stratified by neurological diagnosis. Although radiomics shows great potential for increasing diagnostic precision and improving treatment quality in neurology, several limitations hinder its clinical implementation. Addressing these challenges requires collaborative efforts, advancements in image harmonization methods, and the establishment of reproducible and standardized pipelines with transparent reporting. By overcoming these obstacles, radiomics can significantly impact clinical neurology and enhance patient care.
\end{abstract}

% keywords can be removed
\keywords{Radiomics \and Neuroimaging \and Medical image analysis \and Imaging features \and Radiology \and Deep learning}

\section{Introduction}

Since the discovery of X-rays \cite{Bercovich2018-kt}, medical imaging has advanced significantly. However, the conventional manual review of medical images by clinicians is subjective, time-consuming, and costly. With the increasing availability of medical imaging data, there is a growing opportunity for quantitative analysis in this field. \\
Radiomics is a methodology aimed at retrieving quantitative information from imaging data \cite{lambin2012radiomics}. It is based on extracting numerous descriptors from medical images and finding a link between features and clinical outcomes. Handcrafted radiomics utilizes termed features, which are mathematically defined during the pipeline development, whereas deep radiomics utilizes features created by the artificial neural network during the model training process. The radiomics approach hypothesizes that medical imaging data contains hidden, complementary biological information that can be used for decision support in clinical practice \cite{Guiot2022-zr}. Therefore, this method is of high interest for application in individualized diagnosis and treatment.\\
As handcrafted radiomics workflow requires a segmented region of interest (ROI), this methodology has been extensively developed in the oncological field where tumors and organs are routinely delineated for treatment planning purposes \cite{lambin2012radiomics}. Since its inception,  pioneer studies have revealed the connection between imaging biomarkers and histology \cite{Aerts2014-pl} and have matured to produce externally validated and clinically relevant predictive models \cite{Noortman2023-um, Guevorguian2023-ox}. \\
Whereas in oncology a large amount of segmented imaging data is accumulated mostly for radiotherapy needs, other branches of medicine have collected imaging data and could potentially benefit from the application of radiomics. Thus, it is essential to perform an early diagnosis of neurological diseases since symptoms appear after the disease progresses considerably. Often there are no formal reliable biomarkers, and the diagnosis is based on the regularly reviewed diagnostic criteria \cite{Sharma2023-aq}. Therefore, the differential diagnosis between the diseases and handling the atypical cases might be challenging.\\
Thus, radiomics is an emerging methodology in medical imaging research expanding from oncology to other branches of medicine. However, the review of radiomics in non-oncological neurology is needed to analyze the current state of the art, identify the methodological pitfalls, and suggest possible solutions for the future progress of quantitative clinical neuroimaging. In this review, we present a typical workflow of radiomics analysis regarding neuroimaging. We provide a broad overview of the currently published works stratified by neurological diagnosis. We discuss the current limitations of radiomics in neurology, suggesting potential improvements.

\section{Workflow}
\begin{figure}[h]
    \centering
    \includegraphics[width=\linewidth]{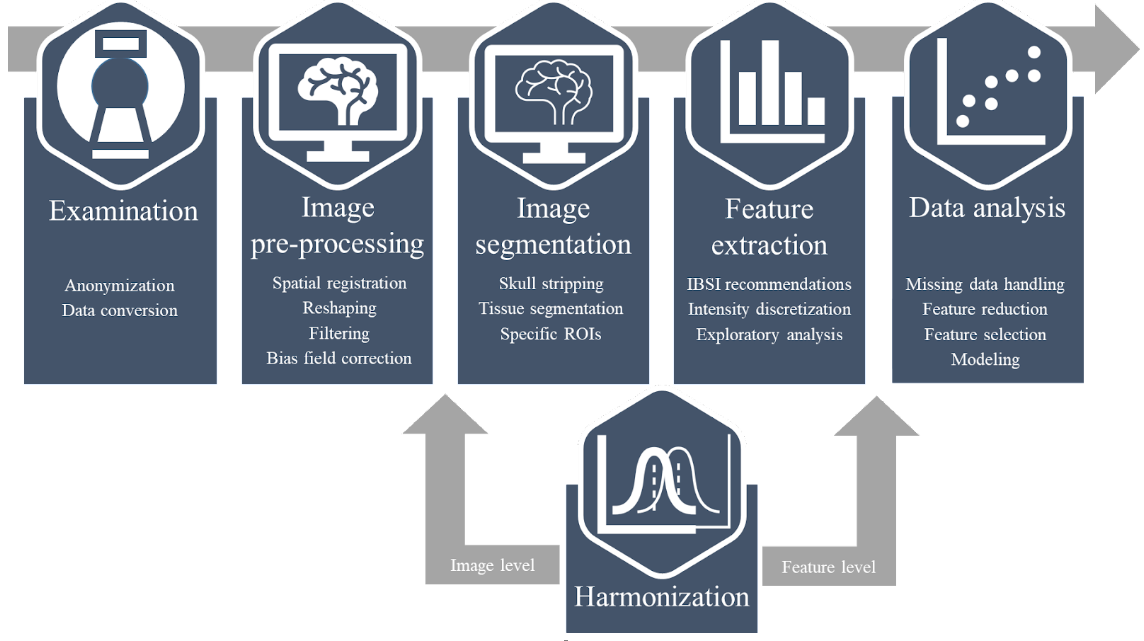}
    \caption{Radiomics pipeline for neuroimaging.}
    \label{fig3-1}
\end{figure}

The following section considers the practical implementation of radiomics in the neuroimaging field combining some common steps of radiomics and neuroimaging workflows (illustrated in \autoref{fig3-1}). After the steps are described, the list of the corresponding software is provided.

\subsection{Data curation}
In hospitals, the data is saved in Picture Archiving and Communications Systems (PACS) in Digital Imaging and Communications in Medicine (DICOM) format \cite{Parisot1995-ds, Rahmim2017-pm}. It stores imaging data together with metadata. In research, open file formats are preferred, such as Nifti, Analyse, MNC, and NRRD \cite{li2016first}. \\
To read and write the imaging and metadata, an application programming interface (API) for the currently relevant programming languages is recommended to make the pipeline fully automated and avoid manually introduced mistakes.\\
Clinical and metadata need to be anonymized or pseudo-anonymized, considering the possible need for follow-up acquisitions \cite{El_Emam2015-nj}. Brain scans usually include the facial features of the patient or teeth. Since facial features or teeth can be used to identify a person, it is necessary to remove them as well. A simple procedure for defacing is skull stripping \cite{Theyers2021-uj}.\\
A good practice is a sanity check of the data. It might include linking imaging and non-imaging samples to reveal missing or unwanted data, acquisition time point check for longitudinal studies and image quality check.\\
In neuroimaging research, the Nifti format is preferred as it is standardized and constructed specifically for neuroimaging data. In the case of data conversion to Nifti with a custom code, it is important to correctly transfer the geometrical parameters of the scan, as described in \url{https://nipy.org/nibabel/coordinate_systems.html}. \\
For automated data analysis, maintaining a uniform data structure is crucial. Data structure for different patients, imaging modalities, and potential acquisition timeframes should be established together with naming conventions for files and folders. For neuroimaging specifically, the Brain Imaging Data Structure (BIDS) \cite{Gorgolewski2016-vf} is recommended. It offers a standardized approach that is suited for multi-modal data and its derivatives and is supported by the community.

\subsection{Data pre-processing}
After the proper curation, data is considered ready for use. The next step is image pre-processing which is described in \cite{Zwanenburg2020-ca}.\\
Since many imaging modalities or image acquisition time points can be combined in neurology, brain scan co-registration is needed. It means that multiple brain scans should be co-aligned to achieve the closest spatial position. The data to be co-registered can belong to different imaging modalities or sequences. Besides co-registration at the patient level, registration to the tissue probability maps in the standardized space can be performed \cite{Dickie2017-gz}. Co-registration can be performed in both rigid (only the head position and orientation are changed) and non-rigid (additional scaling and elastic deformations) ways. Even though non-rigid co-registration allows for the best correspondence of the anatomy and regions of interest, it changes visualized tissue texture. Therefore, for clinical tasks, mostly rigid co-registration on a patient level is applied. \\
Image re-shaping is required to obtain the same voxel shape within the dataset. It allows for the same input image shape in the pipeline. While changing the voxel size, it is important to consider the interpolation effects introduced. In \cite{lehmann1999survey}, different interpolation methods are described. The detailed recommendations are given in \cite{Zwanenburg2020-ca}.\\
Brain scans contain intensity inhomogeneities due to the presence of the bias field. In MRI, the bias field is caused by the MR field inhomogeneity of the scanner originating from the equipment \cite{Padilla2017-pj} and the patient disturbing the magnetic field. To reduce the effect of the bias field, bias field correction (BFC) can be applied. While performing BFC, it is important to consider that it might reduce the contrast and remove critical abnormality information. The most popular method is N4 BFC \cite{tustison2010n4itk}. However, there are recent works on deep learning-based BFC \cite{Chuang2022-np}. CT is not affected by bias because it represents attenuation of the X-ray beam through the body, therefore, in general, this procedure is not recommended for CT scans.\\
Since CT images are expressed in HU, with a well-defined range, more advanced reliable pre-processing is possible. Knowing the characteristic HU for the tissue of interest, it is possible to exclude all the objects on the scan that are not relevant to the analysis. Signal clipping can be applied to the intensities outside of the range of interest \cite{Kandhway2019-xg}.\\
Even though “hard pre-processing” is not recommended in the quantitative image analysis to prevent a signal loss \cite{Zwanenburg2020-ca}, some filtering can be applied to decrease the noise level. The most popular filters among smoothing filters are Gaussian and median filters. Gaussian filter is effective in removing high-frequency noises whereas a median filter is applied to remove impulse noise \cite{kumar2020comparative}. 

\subsection{Image segmentation}
In neurology, ROIs can be anatomically or physiologically derived and vary from application to application. Since brain structures have a systematic organization and traditional computer vision techniques can be applied for segmentation, there are many computer vision-based auto-segmentation tools recognized by the neuroimaging community. Nevertheless, the development of neural networks brings new solutions which are gaining more interest. The deep learning models are trained on different data and do not contain mathematical constraints about anatomy or the expected distribution of intensities.\\
Brain extraction narrows the image size and removes the surrounding tissues. In some studies, radiomics analysis was performed over the whole brain mask \cite{li2022radiomics}. But since the brain includes different structures, whole-brain radiomics do not give comprehensive information about particular shapes and textures. Nevertheless, this kind of analysis is prospective in the discovery of healthy and pathological brain signatures for screening. In some works, features are extracted from the right and left hemispheres to be compared \cite{Ren2022-wy}. \\
A lower level of defining the ROIs is presented with the nervous tissues. The human brain is composed of white and gray matter. Analysis of the radiomics features extracted from the separate brain tissues is closer to the in-vivo histology and allows for the interpretation of texture and density abnormalities. \\
Many studies focus on analyzing certain areas of the cortex and deep gray matter. In this case, it is possible to build connections between imaging data and functional outcomes since particular cortical areas are responsible for the specific functions. Moreover, analyzing certain areas can lead to early disease diagnosis before more severe symptoms appear \cite{Rao2022-dh}. \\
ROIs described above can contain both healthy and pathological tissues. In most cases, neurological pathology appears in textural changes. But when the pathological tissue is compact (hematoma or tissue lesion), it can be selected as ROI.\\
Physiologically-derived ROIs are obtained from functional imaging such as PET or fMRI. They represent the delineated tracer or function activity areas. No manual delineation is needed, and ROI contours depend on the selected binarization method.

\subsection{Feature extraction}
Radiomic features are usually divided into shape-, intensity-, and texture-based features \cite{Aerts2014-pl}. The first category contains mathematical descriptors of ROI geometry, both 2D and 3D. The second category contains intensity statistics and histogram-derived descriptors. The third category contains descriptors of the spatial distribution of image intensity values and their mutual orientation. Shape-based features are not relevant in most neuroimaging studies since the shape of the analyzed ROIs is either standardized or complicated. Nevertheless, volume is always important since it can represent the dystrophy of brain structures, lesion load, or size of the affected area. Intensity- and texture-based features describe tissue properties including tissue homogeneity and density, therefore these groups of features are used in neurological studies.\\
Additional features to the ones obtained from the original image can be extracted when the same values are calculated from the transformed images. Image transformations may include, but are not limited to, square, square root, logarithm, exponential, Gaussian, Laplacian, Laplace of Gaussian, wavelet, local binary pattern, and Gabor filters.\\
While performing feature extraction, besides image and ROI mask, feature extraction parameters are needed. Different feature extraction tools provide different levels of customization. IBSI has some recommendations on the most common feature extraction parameters. This includes intensities re-scaling with normalization or z-scoring. Intensity re-scaling for the images expressed in arbitrary units (MRI) is recommended but it is not for (semi) quantitative images, such as CTs. Another feature extraction parameter is intensity discretization before extracting texture features \cite{Duron2019-cz}. 

\subsection{Data analysis}
After the feature extraction step, data analysis as well as model development and validation are performed. For this, there are a large number of publications on good practices in AI \cite{Kakarmath2020-tc, varoquaux2022machine}. Therefore, in this section, we will focus on radiomics-specific steps.\\
Every case of missing data raises a question of feature or patient elimination, or data imputation. In \cite{Fan2021-uh}, a histological data imputation approach was suggested relying on the present features. Excluding patients will limit the population. Excluding features will limit the amount of diagnostic information. However, crucial clinical or demographic information missing should lead to record exclusion. Since the radiomic features are highly intercorrelated, in case of the absence of a radiomic feature, it can be both eliminated or imputed.\\
While developing a radiomics signature, it is important to assess feature stability and exclude non-reproducible features. To detect reproducible features, test-retest studies should be performed \cite{Van_Timmeren2016-cn, Jha2021-vb}. Additionally, stability does not mean informativity, therefore, further steps on feature selection are needed.\\
After the data is split into train and test sets, the test set should be kept apart and used only in model evaluation so that they do not interfere with the model building process. In some cases, when the data size is not large or to show model robustness to data deviation, cross-validation is performed. This means that the data is split in one of the common cross-validation schemes multiple times and the whole training process is performed from scratch for every split. To show model generalizability, a good practice is to perform external validation —- to demonstrate model performance on external data coming from different acquisition equipment or hospitals. \\
To get rid of the redundant information in the data, inter-correlated features should be excluded preserving the information content. Additionally, it is necessary to exclude features with zero and low variance since they may contain little signal. To detect non-variant features, the standard deviation is usually calculated followed by scaling to the mean value. This approach gives unstable results if feature values have significantly different ranges and mean values. One of the popular methods is implemented as a nearZeroVar function of Caret package (\url{https://topepo.github.io/caret/}) in R. \\
The final step before modeling is feature selection to only retain the informative features. Usually, feature selection steps are model-based. Therefore, for different machine learning models, different features might be selected. The feature set should be reported together with the model performance: if the resulting model performs with low scores, we cannot conclude that the features are strongly linked to the outcome. Feature selection can be performed based on the univariate feature performance \cite{Parmar2015-dy}, feature weights in the model, or “recursive feature elimination” (RFE) based on recursively decreasing the feature set size and comparing model performances \cite{guyon2002gene}. The number of the features in the final feature set can be estimated based on the sample size using several rules of thumb \cite{hua2005optimal, Abu-Mostafa2012-uo}, or empirically based on the saliency point of the dependency of the model score from the number of features.\\
Dimensionality reduction methods such as principal component analysis (PCA) \cite{Anne-Leen2022-zm}, independent component analysis (ICA) \cite{zhang2017functional}, or linear discriminant analysis (LDA) \cite{Fan2019-yq} can also be used to decrease the model complexity. This group of methods is based on the features decomposition resulting in the input matrix transformation into a lower-dimensionality matrix containing only distinctive information. However, it is not commonly used in radiomics since it transforms transparent radiomic features into abstract values losing the interpretability of the final method.\\
In the case of deep radiomics, feature reduction, and selection procedures are performed by the neural network in a data-driven manner.\\
At this stage, overfitting as one of the major problems in machine learning should be considered. Comparison of the training and testing set scores enables assessment of an overfitting effect. A way to examine the model for overfitting is a permutation test \cite{de2020comparison}. To prevent overfitting, several techniques can be applied. These techniques are based on introducing random components into the data or the model and include data augmentation, regularization, ensembling, early stopping, or dropout layer for deep radiomics. 

\subsection{Harmonization}
After the whole radiomics pipeline is established, the next steps are larger multi-center studies or clinical trials. One of the challenges here is feature instability caused by variations in population or acquisition equipment. It leads to situations where models are performing poorly on the data from the unseen domain. Harmonization is data alignment ensuring its compatibility and consistency. As it was shown in \cite{Mali2021-bz}, harmonization can be performed at different steps of the radiomics pipeline, but globally in the image or feature domain.\\
In multi-center studies, data harmonization can start at the beginning of the study by standardizing image acquisition. Nevertheless, even when the protocols are standardized, the intensity distribution in scans can vary, especially while dealing with MRI data. To identify stable features, test-retest and phantom studies can be performed \cite{Jha2021-vb}.\\
If the raw sensor data is available, scans can be reconstructed with the same parameters \cite{Gallardo-Estrella2016-sz}. It is possible to implement traditional image processing methods (intensity normalization, z-scoring) as well as deep-learning-based style transfer. Nevertheless, while changing the appearance of the scans and their intensity distribution, it is not clear whether it will improve feature analysis. In \cite{dewey2019deepharmony}, the U-Net is trained to produce the MRI scans with consistent contrast. In \cite{selim2021ct}, the generative adversarial network was trained to harmonize MRI scans aiming to preserve the consistency of feature values.\\
Finally, feature values can be harmonized. The most popular feature harmonization method is ComBat originally developed for harmonization of the gene expression data \cite{Orlhac2022-xn}. It is an empirical Bayesian method aimed at removing batch-specific bias and preserving the influence of biologically significant components. For the neuroimaging data, DeepCombat combining conditional variational autoencoder architecture with ComBat methodology is suggested \cite{Hu2023-kx}.

\subsection{Handcrafted vs. deep radiomics}
\begin{figure}[h]
    \centering{center}
    \includegraphics[width=\linewidth]{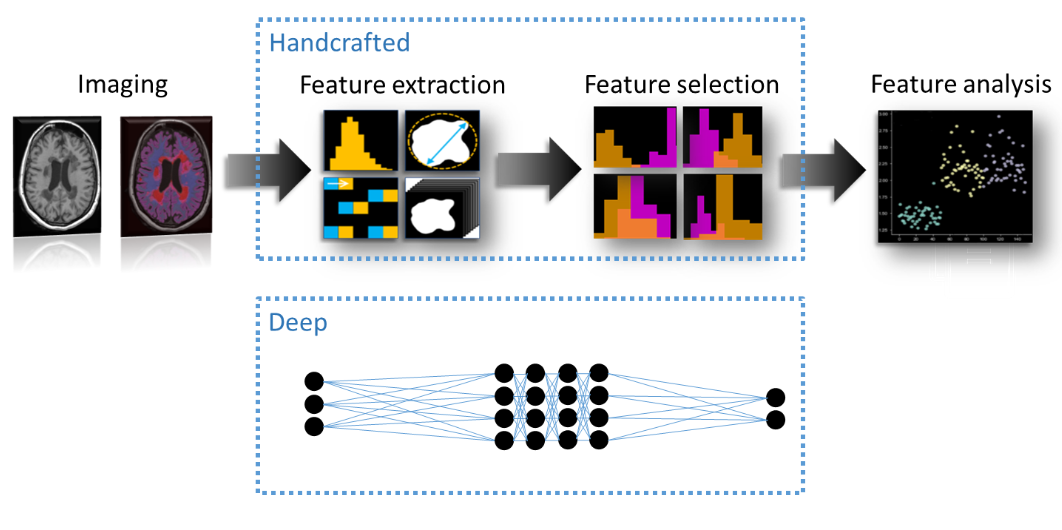}
    \caption{Radiomics pipeline: steps of the handcrafted radiomics represent the "hard-coded" implementation of the deep radiomics.}
    \label{fig3-2}
\end{figure}

Deep radiomics automatically learns representative image features from the high-dimensional data by using non-linear modules of the neural network \cite{Schmidhuber2015-la}. Handcrafted radiomics represents a “hard-coded” version of deep radiomics. Whereas in handcrafted radiomics ROI, feature formulas, and mathematical models are defined by the user, in deep radiomics, these instances are learned from the data. Therefore, feature extraction and selection can be replaced with the neural network. However, a neural network can be a supplementary component for the handcrafted workflow to segment the ROI. In the most general case, the whole scan can be used as a neural network input, as illustrated in \autoref{fig3-2}. In this case, the model will learn to identify the informative features of the scan.\\
Both handcrafted and deep radiomics have advantages and limitations. Handcrafted radiomics can be trained with less data and is more transparent due to the interpretable features. However, it is limited in capturing complex patterns, not robust to variations of imaging parameters, and requires image preprocessing and segmentation. Deep radiomics learns relevant features automatically, captures more complex dependencies, and shows impressive results. It can adapt to the different imaging modalities and tasks with minimal changes in architecture. Moreover, additional domain knowledge can be utilized with transfer learning. However, deep radiomics is greedy for training data and computational resources and is challenging to interpret \cite{rogers2020radiomics, Wagner2021-vz}.\\
Not all of the handcrafted features are necessarily linked to the outcome. In contrast, deep radiomics generates features during the training process. Recently attention in AI has shifted towards self-supervised learning and foundation models \cite{Krishnan2022-id, moor2023foundation}. Self-supervised learning aims to provide pseudo-labels to the data by deriving supervisory signals from the data. After pre-training, models can be fine-tuned for specific downstream tasks. This approach allows for training with a substantial amount of unlabeled data and enables utilization of the same pseudo-labels for multiple downstream tasks creating a new concept of imaging features.

\subsection{Neuroimaging software packages}
To sum up, the most acknowledged open-source neuroimaging software packages are presented in \autoref{table3-1}. At the moment, the most common language for open-source research software is Python. \autoref{table3-2} presents Python packages for neuroimaging. 

    \begin{table}[t]
\caption{Neuroimaging software.}
\label{table3-1}
\resizebox{500pt}{!}{%
\begin{tabular}{llll}
\hline
Name       & Link                                                 & Platform/environment & Functionality                  \\ \hline
3D Slicer \cite{Kikinis2014-al} &
  \url{https://www.slicer.org/} &
  Desktop, API &
  \begin{tabular}[c]{@{}l@{}}Images visualization, processing, \\ segmentation,   registration, \\ and analysis; planning and navigating \\ image-guided procedures\end{tabular} \\
ITK-Snap \cite{mccormick2014itk}   & \url{http://www.itksnap.org/}                              & Desktop, API         & Image segmentation             \\
RadiAnt    & \url{https://www.radiantviewer.com/}                       & Desktop              & Viewer                         \\
MicroDicom & \url{https://www.microdicom.com/}                          & Desktop              & Viewer                         \\
SPM12 \cite{Penny2011-gd}      & \url{https://www.fil.ion.ucl.ac.uk/spm/software/spm12/}    & MATLAB               & Complete analysis package      \\
Dcm2nii    & \url{https://www.nitrc.org/projects/dcm2nii/}              & Desktop, CLI         & DICOM to Nifti                 \\
Dcm2niix   & \url{https://github.com/rordenlab/dcm2niix}                & CLI                  & DICOM to Nifti conversion      \\
Dcm2niir   & \url{https://github.com/muschellij2/dcm2niir}              & R                    & DICOM to Nifti conversion      \\
FSL \cite{jenkinson2012fsl}        & \url{https://fsl.fmrib.ox.ac.uk/fsl/fslwiki}               & Desktop, CLI, API    & Complete analysis package      \\
Freesurfer \cite{dale1999cortical} & \url{https://surfer.nmr.mgh.harvard.edu/}                  & Desktop, CLI, API    & Complete analysis package      \\
ANTs \cite{avants2011reproducible} &
  \url{https://github.com/ANTsX/ANTs} &
  Unix scripting &
  \begin{tabular}[c]{@{}l@{}}Complete analysis package \\ (dependent on fMRI packages)\end{tabular} \\
Brainsuit \cite{shattuck2002brainsuite}  & \url{https://brainsuite.org/}                              & MATLAB               & Complete analysis package      \\
LST \cite{schmidt2012automated}        & \url{https://www.applied-statistics.de/lst.html}           & MATLAB + SPM         & WM lesion segmentation         \\
Camino \cite{basser1994estimation}     & \url{http://camino.cs.ucl.ac.uk & CLI}                  & Diffusion MRI processing       \\
QuickNAT \cite{roy2019quicknat}   & \url{https://github.com/ai-med/QuickNATv2}                 & MATLAB               & Neuroanatomy segmentation      \\
AFNI \cite{cox1996afni} &
  \url{https://afni.nimh.nih.gov/} &
  Desktop, CLI, API &
  Complete analysis package \\
CAT \cite{gaser2022cat}        & \url{https://neuro-jena.github.io/cat//}                   & MATLAB + SPM         & Computational anatomy analysis \\ \hline
\end{tabular}%
}
\end{table}

\begin{table}[t]
\caption{Neuroimaging Python packages.}
\label{table3-2}
\resizebox{500pt}{!}{%
\begin{tabular}{lll}
\hline
Name        & Link                                       & Functionality                                           \\ \hline
Pydicom \cite{Mason2011-kt}     & \url{https://pydicom.github.io/}                 & Reading, modifying and writing DICOM data               \\
Medpy \cite{oskar_maier_2019_2565940}       & \url{https://github.com/loli/medpy}              & Medical image manipulation                              \\
Nibabel \cite{Brett2023-od} &
  \url{https://nipy.org/nibabel/} &
  \begin{tabular}[c]{@{}l@{}}Reading, modifying and writing common \\ neuroimaging data formats\end{tabular} \\
Pydeface \cite{omer_faruk_gulban_2022_6856482}    & \url{https://github.com/poldracklab/pydeface}    & Defacing                                                \\
Mridefacer  & \url{https://github.com/mih/mridefacer}          & Defacing                                                \\
Mriqc \cite{esteban2017mriqc}       & \url{https://github.com/nipreps/mriqc}           & Image quality assessment                                \\
Miqa        & \url{https://github.com/OpenImaging/miqa}        & Image quality assessment                                \\
ImageQC     & \url{https://github.com/EllenWasbo/ImageQC}      & Image quality assessment                                \\
\begin{tabular}[c]{@{}l@{}l@{}}Precision \\ medicine \\ toolbox \cite{lavrova2023precision}\end{tabular} &
  \url{https://github.com/primakov/precision-medicine-toolbox} &
  \begin{tabular}[c]{@{}l@{}}Image format conversion, metadata collection, \\ basic quality check and pre-processing, \\ basic exploratory analysis of the tabular data\end{tabular} \\
Pybids \cite{yarkoni2019pybids}      & \url{https://github.com/bids-standard/pybids}    & BIDS data   management                                  \\
Deepbrain   & \url{https://github.com/iitzco/deepbrain}        & Brain extraction                                        \\
DeepBleed \cite{sharrock20213d}   & \url{https://github.com/msharrock/deepbleed}     & Haemorrhage segmentation                                \\
Pyradiomics \cite{van2017computational} & \url{https://github.com/AIM-Harvard/pyradiomics} & Radiomic features extraction                            \\
Nipype      & \url{https://github.com/nipy/nipype}             & Pipeline implementation \\
TorchiO \cite{perez2021torchio} &
  \url{https://torchio.readthedocs.io/} &
  \begin{tabular}[c]{@{}l@{}}Medical image augmentation and transformations \\ including noise and artefacts generation\end{tabular} \\
HippoDeep \cite{zavaliangos2022testing}   & \url{https://github.com/bthyreau/hippodeep}      & Hippocampus segmentation                                \\ \hline
\end{tabular}%
}
\end{table}

\section{Applications}
\begin{figure}[h]
    \centering{center}
    \includegraphics[width=\linewidth]{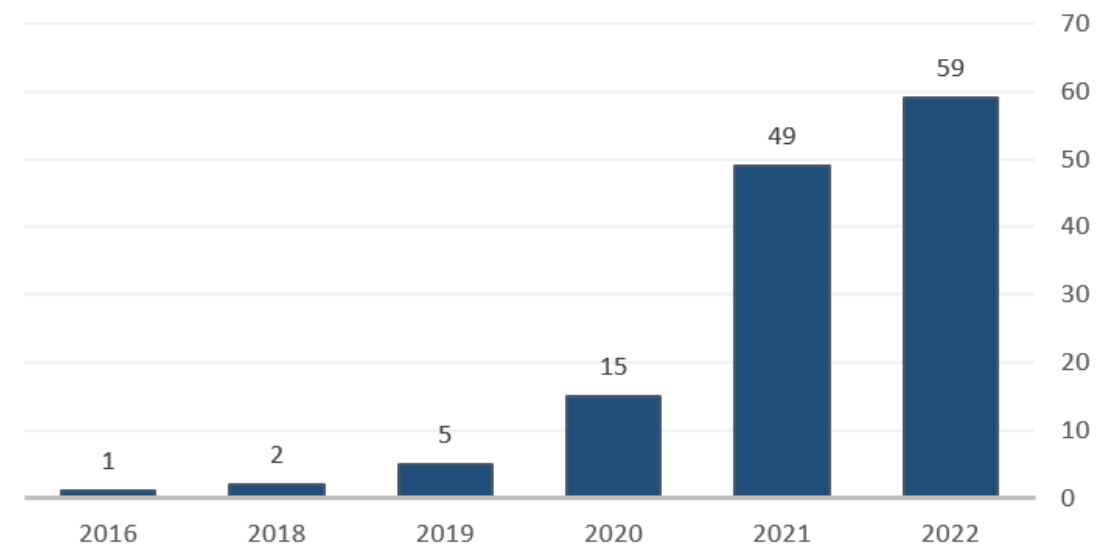}
    \caption{Number of publications, by year, containing keywords ‘radiomics AND neurology NOT oncology’ in PubMed database (\url{https://pubmed.ncbi.nlm.nih.gov/term=radiomics+AND+neurology+NOT+oncology}).}
    \label{fig3-3}
\end{figure}

The number of papers in non-oncological neurology is growing starting from the first publication in 2017 \cite{Rahmim2017-pm}, which explicitly mentions radiomics. \autoref{fig3-3} shows the number of papers per year found in the PubMed database with the search query ‘radiomics AND neurology NOT oncology’. In this section, we will review the non-oncological neurological studies utilising radiomics.

\subsection{Alzheimer's disease}
Alzheimer’s disease (AD) is the leading cause of dementia worldwide, and dementia is the leading cause of disability among the elderly population \cite{who2023dementia}. Since research is extensively performed in this field, and much experience and data have been accumulated, the first neurological radiomic studies were performed in this area. \\
As the disease develops gradually, causing different levels of disability, the first work \cite{Liu2015-ts} classified disease stages with deep radiomics using data from Alzheimer’s Disease Neuroimaging Initiative (ADNI) database \cite{Petersen2010-ku}. This first study pointed out the need for more data and for establishing the connection between the radiomic features and pathological processes. There are some later works on AD stages classification as well \cite{Ramzan2019-iu}. Many later works performed binary classification between AD patients and normal controls (NC) to show that MRI- and PET-derived features can be associated with AD \cite{chaddad2018deep, Qiu2018-jg, Zhao2020-dj, Qiu2020-lh, Tufail2020-oa, Jo2020-tc, Duc2020-na, Liu2022-we, du2023deep}. Whereas in AD, pathological changes might be well visible in scans and be accompanied by strong clinical symptoms, mild cognitive impairment (MCI) is harder to diagnose. Moreover, MCI individuals can be confused with AD patients and NC. There are works on distinguishing between NC and MCI patients \cite{Feng2019-hl, Ju2019-mq}. Most of the AD radiomics studies are performed on classification between NC, MCI, and AD patients utilizing either pair-wise binary or multi-class classification \cite{Shi2018-fg, Ding2019-rf, feng2018radiomic, Amoroso2018-vl, Li2019-iq, Ranjbar2019-yv, Basaia2019-bg, Pan2020-vu, Ding2021-gp, Feng2021-mq, Du2021-nx, Zhao2022-up}. As MCI is considered as an early stage of AD, for these patients it is crucial to know whether their impairment will progress to AD. In \cite{Kim2021-se}, the classification model is trained to estimate amyloid positivity status in MCI patients. In a number of works, classification models are built to directly predict the conversion from MCI to AD \cite{Huang2020-zk, Zhou2018-hr, Spasov2019-jv, Lee2019-tn, Li2020-la, Abrol2020-er, Yang2022-ne}. There were also attempts to predict the speed of disease progression \cite{Li2018-aj}. Nevertheless, AD is not the only cause of dementia, and focusing on AD patients only will lead to the lack of specificity of the methods. There are works on the classification of the different dementia diagnoses: dementia with Lewy bodies vs AD \cite{Li2018-aj, Iizuka2019-cd} and idiopathic normal pressure hydrocephalus vs. AD \cite{Irie2020-tx}. \\
Nevertheless, besides clinical radiomics, there are traditional neuroimaging features to characterize the brain. Functional connectivity plays an important role here. Even though a number of studies relied on fMRI \cite{Feng2019-ho, Ju2019-mq, Ramzan2019-iu, Li2020-la, Duc2020-na}, the connection between these biomarkers has yet to be revealed. In \cite{Feng2019-ho}, the correlation between connectivity and radiomic features is studied.

\subsection{Multiple sclerosis}
Multiple sclerosis (MS) is the leading cause of disability among the young population \cite{lassmann2018multiple}. The disease progression is fast, therefore early diagnosis and prognosis for the patient are important. \\
However, the first deep radiomic study on inflammatory degeneration was performed on survival classification of the amyotrophic lateral sclerosis patients \cite{Van_der_Burgh2017-kc}. MS-related studies were performed later and covered simple binary classification between MS and NC \cite{yoo2018deep, Eitel2019-rj, lavrova2021exploratory}. These works show the utility of the approach but lack the specificity of MS among other inflammatory neurodegenerative diseases. Neuromyelitis optica-spectrum disorder (NOSD) can be easily confused with MS therefore a precise diagnosis is needed to enable the correct treatment. There are some works on binary classification between MS and NOSD \cite{Liu2019-lp, ma2019quantitative, Wang2020-tp, Hagiwara2021-vi}. There are studies where the classification models are built to distinguish between MS and other diagnoses such as neuropsychiatric systemic lupus erythematosus \cite{Luo2022-bu} and ischemic vasculopathy \cite{He2022-wl}. The multi-class classifier was built to distinguish between MS, NOSD, migraine, and vasculitis \cite{Rocca2021-ay}. Besides diagnosis, it is essential to grade the disease severity within the MS cohort. In \cite{Du2023-zj}, the radiomics model was trained to estimate the relapse rate in MS. In \cite{roca2020artificial}, EDSS score was predicted with deep radiomics. In \cite{Zhang2021-ko}, MS types were classified as well as NC. \\
Several models are trained to characterize the neurodegeneration process. In \cite{Wei2020-te}, the deep learning model is built to perform the detection of the demyelinated voxels on PET. In \cite{Ye2020-em}, MS lesion classification is performed. In \cite{Barquero2020-vo, Zhang2022-rp}, lesion rim status classification is performed. In \cite{Peng2021-lu}, a handcrafted radiomic model was built to predict lesion growth. 

\subsection{Parkinson's disease}
Parkinson’s disease (PD) is a degenerative condition of the central nervous system, which develops disability faster than any other neurological disorder \cite{who2023parkinson}. Since it develops gradually, exploration of the early and specific biomarkers is essential.\\
The first work was on the motor function assessment in PD \cite{Rahmim2017-pm}. To show the utility of radiomics, the models were built to distinguish between PD patients and NC \cite{Cheng2019-tw, Wu2019-xa, Xiao2019-yy, Cao2020-rm, Liu2020-wx, Shu2020-kg, Chakraborty2020-pv, Cao2021-nd, Yasaka2021-qc, Tafuri2022-mn, Shiiba2022-am, Sun2022-wd}. As PD is heterogeneous in terms of its clinical phenotype, for treatment and severity estimation, PD variants classification is important. The models to classify the patients between parkinsonism subtypes were built in a number of works \cite{Zhao2019-jh, Kiryu2019-yo, Pang2020-jm, Sun2021-cx, Jankovic2021-hn, Shi2022-qz, Pang2022-qq, Zhao2022-qx}. In PD there is a need for differentiation from other similar diseases. In \cite{Hu2021-pn, Tupe-Waghmare2021-sj}, binary classification models were trained to distinguish between PD and multiple system atrophy and progressive supranuclear palsy. Radiomics approach was also used for the treatment response prediction \cite{Shin2021-ue}.\\
Transfer learning attempts are performed in the PD field trying to fine-tune the AD diagnostic model for PD diagnosis \cite{Choi2020-bl}. Diagnostic support solutions are suggested as in \cite{Shin2021-ue}, where the nigrosome 1 abnormalities detector is trained. 

\subsection{Stroke}
Stroke is the second leading cause of death, and third leading cause of disability worldwide \cite{who2023stroke}. The main task in stroke management is patient outcome prediction. \\
As a first task in stroke management, stroke areas should be identified. In \cite{Nishio2020-sa}, the deep learning model was trained to detect the stroke area on non-contrast CT scans. In \cite{Guo2022-rk}, the model was trained to distinguish between hyperperfusion areas from normal ones. In \cite{Lyu2023-rx}, primary and secondary hemorrhages were classified. The other works were devoted to the prognosis of stroke area development \cite{Nielsen2018-yd, Chen2021-yu, Xie2022-ba, Song2023-dn}. Besides the pathology development prediction, there were studies focused on the prediction of the biological recovery processes. In \cite{Aktar2020-cj}, collateral circulation was classified. In \cite{Qiu2019-bn, Hofmeister2020-rc}, models were trained to predict recanalization. Most works were focused on the treatment outcome prediction, including functional and cognitive, for both thrombolysis and mechanical thrombectomy \cite{Hilbert2019-pz, Chauhan2019-qk, Nishi2020-qs, Bacchi2020-pa, Li2022-ul, Tolhuisen2022-vp}. There are studies on stroke onset time estimation with radiomics \cite{Zhang2022-ox}. Finally, some studies are performed to train the radiomic models to predict post-stroke events such as recurrent stroke or epilepsy \cite{Li2022-dj, Wang2022-sh, Lin2023-vt}.

\subsection{Epilepsy}
Epilepsy is the most common neurological disease \cite{who2023epilepsy}. In \cite{Cheong2021-sz}, the model was trained to predict epilepsy laterality. In \cite{Zhang2021-nd}, the epileptic foci detection method was suggested. In \cite{Feng2021-qu}, the radiomic model was trained to detect focal cortical dysplasia lesions. In \cite{Kim2022-jk}, the binary classifier was built to distinguish between Juvenile Myoclonic Epilepsy and NC. 

\subsection{Mental disorders}
Mental disorders affect behavior and quality of life significantly and are observed in 12.5\% of the population \cite{who2023mental}.  Most studies in radiomics in mental disorders are devoted to the binary classification between the NC and schizophrenia patients \cite{kim2016deep, Cui2018-pa, Zeng2018-ka, Yan2019-dx, Qureshi2019-bz, Park2020-rz, Li2020-sk, Oh2020-zs, Hu2022-kw}, bipolar disorder patients \cite{Wang2020-tn}, or first episode psychosis \cite{Vieira2019-jo, zhao2020functional, li2021deep}. In \cite{zhao2020functional}, a more difficult deep radiomics classifier was built to distinguish between schizophrenia patients, major depressive disorder, and NC. In \cite{li2021deep}, the classifier was built to distinguish between first-episode psychosis, bipolar disorder, and NC. Finally, there are some works on radiomics implementation for treatment response prediction \cite{Cui2021-jy, Cui2021-lc}.

\subsection{Neurodevelopmental disorders}
The most common neurodevelopmental disorders are autism spectrum disorder (ASD) and attention-deficit/hyperactivity disorder (ADHD). They affect cognitive and behavioral functions and might require life-long care and support. In most radiomic studies, binary classification models were built to distinguish between NC and ASD \cite{Chaddad2017-ju}, and ADHD \cite{Sun2017-aq, Liu2023-si}. In \cite{Chaddad2017-do}, ASD-linked radiomic features were revealed.

\subsection{Open-source datasets}
We believe that to enable extensive research in some diagnostic areas it is essential to have access to a sufficient amount of medical imaging data. \autoref{table3-3} presents some popular open medical imaging datasets containing neuroimaging data.

\begin{table}[]
\caption{Some of the open-source neuroimaging datasets.}
\label{table3-3}
\centering
\begin{tabular}{lll}
\hline
Name                 & Modality                     & Cohort \\ \hline
ADNI \cite{Petersen2010-ku}                 & MRI, PET                     & AD     \\
OASIS \cite{Chen2020-pk}                & MRI, PET, CT                 & NC, AD \\
ATLAS \cite{Liew2021-te}                & T1w MRI                      & Stroke \\
IXI Dataset \cite{ixidataset}          & T1w, T2w, PDw, MRA, dMRI     & NC     \\
Yale Test-Retest Data \cite{Noble2017-lg} &
  Anatomical and functional MRI &
  NC \\
ISLES \cite{Hakim2021-zz} &
  MRI, CT &
  Stroke \\
MS Lesion Segmentation Challenge 2015 \cite{carass2017longitudinal} &
  T1w, T2w, FLAIR, PDw MRI &
  MS \\
NFBS skull stripped repository \cite{Puccio2016-iu} &
  T1w MRI &
  Psychiatry \\
Calgary-Campinas-359 \cite{souza2018open} & T1w MRI                      & NC     \\
CT-ICH \cite{Hssayeni2020-be}               & CT                           & TBI    \\
RSNA Intracranial Hemorrhage \cite{rsnaintracranial} &
  CT &
  Stroke \\
PPMI \cite{marek2011parkinson}                 & Clinical and biological data & PD     \\ \hline
\end{tabular}
\end{table}

\section{Discussion}
In this review, we gave an overview of the radiomics pipeline in clinical non-oncological neuroimaging and gave some recommendations on each step of the pipeline. The variety of open-source tools for neuroimaging analysis as well as the expanding amount of radiomics studies are bringing optimism in the development of artificial intelligence (AI) in clinical neurology. Moreover, the amount of accumulated data is still growing, pushing the quantitative neuroimaging development forward. Therefore, the approaches, that previously existed in oncology, can be landed in neurology. However, there are some limitations present in the majority of the papers and characterizing the current challenges in the field.

\subsection{Data availability}
Many current studies are cross-sectional and performed on small private datasets. These datasets usually are not sufficient to demonstrate pathological patterns and represent the target cohort. One-center training severely decreases the generalizability of the model. To enable clinical implementation of the model, it needs to confirm its performance on various demographics as well as acquisition equipment set-ups.\\
External validation on samples coming from the different data domains is recommended. It is desirable to have external validation data to be prospectively collected to show robustness to the potential data drift. However, external validation results should be interpreted carefully. Sample size, heterogeneity, and data balance should be considered. External validation performance should be explainable considering the sample properties. The external validation does not give absolute information about model generalizability. There are some industry-inspired suggestions to perform regular and recurrent validation every time the model is deployed to evaluate the generalizability of the predictive model \cite{feng2022clinical, youssef2023all}.\\
Another consequence of the limited data accessibility is the low reproducibility of the published studies. Experiments on private data cannot be repeated. Additionally, it is not possible to check the labeling correctness. Moreover, if one wants to compare some models, he needs to test them on exactly the same cohort. Nevertheless, there are established datasets used by multiple research groups. These are 1) open-source datasets, such as ADNI, 2) challenge datasets (\url{https://grand-challenge.org/}) which are much smaller and usually do not have extensive multi-modal data, 3) clinical trials datasets (for example, \cite{Berkhemer2015-lj}). However, using a single dataset without any external data introduces overfitting across the community, which limits the usefulness of the dataset itself. To make the data and data-driven solutions sustainable and therefore trustful, the following four principles have to be maintained: Findability, Accessibility, Interoperability, and Reusability (FAIR) \cite{wilkinson2016fair}.\\
In the last years, attention has been drawn to the latest generations of AI models – visual transformers and foundation models. They have a huge potential in medicine. Visual transformers demonstrate high performance while solving visual tasks such as image classification or segmentation \cite{dosovitskiy2020image, yang2021t}. Foundation models will be able to solve complex problems on multi-modal data with a high variety of particular downstream tasks \cite{moor2023foundation, Zhang2023-ik}. However, the main challenge in their implementation is the high demand for the amount of the training data. This fact gives an additional motivation for data sharing and aggregation.

\subsection{Data harmonization}
Another limitation resulting from limited data availability is the lack of data harmonization. It is affecting the radiomic methods themselves because of the heterogeneity of the population as well as acquisition equipment. If the development data is not population and equipment representative, every new inference data point can be out of the distribution which leads to the wrong model outcomes. The CT data is presented in HU which gives it quantification and stability. MRI data is expressed in arbitrary units and acquired with a large variety of MR sequences and hardware. One of the promising directions in the MRI research is in multi-echo qMRI maps reconstruction. It gives stability and quantification to the MRI data, but at the moment this approach is far from being used in the clinical set-up. Another approach is in implementation of the AI-based methods such as generative adversarial networks \cite{selim2021ct} to harmonize the data.    

\subsection{Clinical relevance of the data}
Reliable and stable imaging biomarkers for neurological disorders should be not only sensitive but also specific for every neurological condition. In the current studies, most of the models are developed to distinguish between the disorder and NC. Therefore, these approaches are not applicable in clinical practice, where more than one neurological condition can be suspected. Moreover, co-existing conditions are also possible. Therefore, broader studies and intra-diagnosis tests are needed to develop disease-specific radiomic signatures. \\
Most neurological diseases develop gradually, and the patients are diagnosed in already chronic stages of the disease. This brings bias to the study data that almost does not contain non-symptomatic patients at early stages. These patients are highly important to develop methods for early diagnosis and disease prevention. However, there are some longitudinal studies (for example, ADNI-based) that are performed with the early-stage cases.

\subsection{Study design}
Current studies are mostly proof-of-concept. Therefore, the study design is highly simplified. Prediction tasks are solved as classification tasks in most cases and the outcomes are limited by the predicted event timeframe. \\
Imaging modality should be selected correctly, based on domain knowledge, existing clinical protocols, and its availability.\\
The fusion of data of different natures should not be performed before the predictive power of every baseline data source is studied. However, data fusion can be justified by showing the added value in model performance obtained with the fused data compared to the separate baseline models. \\
For handcrafted radiomics, ROI selection should be based on domain knowledge. In the currently published papers, this justified approach was demonstrated, and ROIs were selected based on the brain areas affected the most by the corresponding diseases.

\subsection{Pipeline implementation}
Different research groups perform radiomic pipeline steps differently and in different order. This results in inconsistency and low reproducibility of the results. To overcome this issue, transparent reporting is essential following TRIPOD \cite{Collins2015-ph} and RQS \cite{lambin2017radiomics}. Recently, the CheckList for EvaluAtion of Radiomics Research (CLEAR) checklist for radiomics was out \cite{kocak2023checklist}. \\
Additionally, code sharing has become a common practice in scientific reporting in the last few years making the results transparent and reproducible \cite{hunter2021ten}.\\
Another challenge is caused by the fact that multiple data processing tools are implemented in different platforms and environments breaking the consistency of the data flow. However, for every common neuroimaging tool multiple APIs exist enabling a single infrastructure for the study implementation. \\
To justify the selection of the model, its design, and hyperparameters in a reproducible environment, experiment tracking tools such as MLFlow (\url{www.mlflow.org}) are useful. They do not only inform the researcher about the best-performing setup but also the protocol of all the experiments. \\
Another implementation challenge is related to image segmentation which is traditionally performed manually. Since intra- and inter-reader agreement is never absolute, development and improvement of the automated segmentation methods is needed.\\
For the clinical application of AI in medical imaging, FUTURE-AI guiding principles are developed \cite{Lekadir2021-rf} to ensure that AI solutions are effective, trustworthy, ethical, and safe.

\subsection{Interpretation}
Even though the reported models perform with high scores, there is still a lack of interpretation. For this, behavioral analysis should be performed and aligned with the clinical knowledge. Connections between the predictive and stable radiomic features and clinical parameters should be studied. Additionally, pathological mechanisms are not revealed by the radiomic studies, and large work should be done in this field supported by extensive clinical and histological data. \\
Since medical imaging analysis involves high-stakes decisions, information is needed about which influence inputs have on a final decision of the model. Due to a simple implementation, handcrafted radiomics is more transparent compared to deep radiomics. However, for acceptance in clinical practice, implementation of explainable AI (XAI) is needed for overcoming a "black box problem" \cite{VANDERVELDEN2022102470, SALAHUDDIN2022105111}. \\
While every AI model is accompanied by its performance scores, which provide insights into its efficiency and facilitate comparisons with other models \cite{maier2022metrics}, it is imperative to remember that the significance lies not in the AI scores themselves but in the impact on clinical outcomes. Consequently, for more advanced models, the inclusion of supplementary metrics is vital to elucidate how they enhance the clinical pipeline.

\section{Conclusion}
We gave a review of the radiomic pipeline from the clinical neuroimaging perspective. The amount of collected data and the high performance of the published models have shown that the application of radiomics in neuroimaging will increase diagnostic precision and quality of treatment. Development of the new AI methods will increase the performance of the deep radiomics as well as will enable solutions for more complex and multi-modal tasks. However, some important limitations are preventing the implementation of this methodology in clinical practice for high-level diagnostic support rather than image quantification. To overcome these limitations, it is necessary to set data exchange and collaborations, work on data harmonization methods, and implement reproducible pipelines and transparent reporting.

 \bibliographystyle{elsarticle-num} 
 \bibliography{bibliography}





% \end{thebibliography}
\end{document}